\documentclass[12pt]{article}

\usepackage{amsmath,amsthm, amsfonts, amssymb, amsxtra,amsopn}
\usepackage{mathdots}
\usepackage{graphicx}
\usepackage{multirow}
\usepackage{algorithmic}
\usepackage{longtable}
\usepackage{booktabs}
\usepackage{listings}
\usepackage{morefloats}
\usepackage{cmap}
\usepackage{tikz}
\usepackage{pgfplots}
\usepackage{pgfplotstable}
\pgfplotsset{compat=1.7}
\usepackage{tabularx}
\usepackage{hyperref}
\hypersetup{colorlinks=true,linkcolor=black,citecolor=black,urlcolor=black,filecolor=black}
\hypersetup{pdfpagemode=UseNone,pdfstartview=}

\advance\oddsidemargin by -0.6in
\advance\textwidth by 1.2in

\advance\topmargin by -0.36in
\advance\textheight by 0.72in

\def\eref#1{(\ref{#1})}

\def\O{{\cal O}}
\def\z{\phantom{z}}

\def\TP{\hbox{TP}}
\def\FP{\hbox{FP}}
\def\TN{\hbox{TN}}
\def\FN{\hbox{FN}}
\def\TPR{\hbox{TPR}}

\def\FPR{\hbox{FPR}}

\def\rec{\hbox{recall}}
\def\prec{\hbox{precision}}

\long\def\symbolfootnotetext[#1]#2{\begingroup%
\def\thefootnote{\fnsymbol{footnote}}\footnotetext[#1]{#2}\endgroup}

\title{A Comparison of Static, Dynamic, and Hybrid Analysis for Malware Detection}

\author{Anusha Damodaran\footnotemark[1]\ \ \ 
Fabio Di Troia\footnotemark[2]\ \ 
Visaggio Aaron Corrado\footnotemark[2]\\
Thomas H.~Austin\footnotemark[1]\ \ 
Mark Stamp\footnotemark[1]\,\,\,\footnotemark[3]}

\begin{document}

\symbolfootnotetext[1]{Department of Computer Science, San Jose State University}
\symbolfootnotetext[2]{Department of Engineering, Universit\`{a} degli Studi del Sannio}
\symbolfootnotetext[3]{stamp$@$cs.sjsu.edu}

\maketitle

%
%
\abstract

In this research, we compare malware detection techniques based on
static, dynamic, and hybrid analysis. Specifically,
we train Hidden Markov Models (HMMs ) on both static and dynamic feature sets
and compare the resulting detection rates over a substantial number of malware families.
We also consider hybrid cases, where dynamic analysis is used in 
the training phase, with static techniques used in the detection phase, and vice versa. 
In our experiments, a fully dynamic approach generally yields the best detection rates.
We discuss the implications of this research for malware detection 
based on hybrid techniques.

%
%
\section{Introduction\label{sect:introduction}}

According to Symantec~\cite{symantecreport}, more than~317 million new 
pieces of malware were created in~2014, which represents a~26\%\ increase
over~2013. Given numbers such as these, malware detection is clearly
a significant and worthwhile research topic.

In practice, the most widely used malware detection method is signature 
scanning, which relies on pattern matching.
While signature scanning is effective for many types of malware,
it is ineffective for detecting new malware, or even significant variants of existing 
malware~\cite{aycock}. 

A wide array of advanced detection techniques have been considered in the literature. 
Some detection techniques rely only on static 
analysis~\cite{chinmayee,Attaluri,BLS,mihai,pdeshpande,DPS,JAS,shankarpani,Shanmugam,Tanuvir,sorokin}, that is, features that
can be obtained without executing the software. In addition, dynamic analysis
has been successfully applied to the malware detection 
problem~\cite{ahmed,Anderson,ratan,mojtaba1,hdm2,kolbitsch,park,youghee,yong,rhee,imds}. 
Recently,
hybrid approaches have been analyzed, where both static and
dynamic features are used~\cite{young,mojtabahdm}.

Here, we compare static analysis with dynamic analysis, and also consider
hybrid schemes that combine elements of both. We use a  
straightforward training and scoring 
technique based on Hidden Markov Models and we consider
feature sets consisting of API call sequences and opcode
sequences. 

In this research, our goal is to gain some level of understanding of
the relative advantages and disadvantages of static, dynamic,
and hybrid techniques. In particular, we would like to determine whether there is
any inherent advantage to a hybrid approach. Note that our goal here is not to
optimize the detection accuracy, which would likely require combining a variety of scores
and scoring techniques. 
Instead, we conduct our analysis in
a relatively simple setting, by which we hope to reduce the number of potentially confounding
variables that tend to appear in more highly optimized systems.



The remainder of this paper is organized as follows.
In Section~\ref{sect:background}, we discuss relevant 
background information, including related work.
Section~\ref{sect:implementation} discusses the experiments conducted
and the datasets used. 
In Section~\ref{sect:experiments}, we present our experimental results.
The paper concludes with Section~\ref{sect:conclusion}, where we also
mention possible future work. 

\section{Background\label{sect:background}}

In this section, we first provide a brief discussion of 
malware detection techniques,
with an emphasis on Hidden Markov Models, which are 
the basis for the research presented in this paper. 
We also review relevant related work.
Finally, we discuss ROC curves, which give us a convenient
means to quantify the various experiments that we have conducted.

\subsection{Malware Detection}

There are many approaches to the malware detection problem. Here, we briefly consider
signature-based, behavior-based, and statistical-based detection,
before turning our attention to a slightly more detailed discussion of HMMs.

\subsubsection{Signature Based Detection}

Signature based detection is the most widely used anti-virus technique~\cite{aycock}. 
A signature is a sequence of bytes that can be used to identify specific malware.  
A variety of pattern matching schemes are used to scan for signatures~\cite{aycock}.
Signature based anti-virus software must maintain a repository of signatures of known malware
and such a repository must be updated frequently as new threats are discovered. 

Signature based detection is simple, relatively fast, and effective against most
common types malware. A drawback of signature detection is that it requires 
an up-to-date signature database---malware not present in the database will not be detected. 
Also, relatively simple obfuscation techniques can be used to evade signature
detection~\cite{Me}.

\subsubsection{Behavior Based Detection}

Behavior based detection focuses on the actions performed by the malware during 
execution. In behavior based systems, the behavior of the malware and benign files are 
analyzed during a training (learning) phase.
Then during a testing (monitoring) phase, 
an executable is classified as either malware or benign, based on
patterns derived in the training phase~\cite{behavsurvey}.

\subsubsection{Statistical Based Detection}

Malware detection can be based on statistical properties derived from program features. 
For example, in~\cite{wong}, Hidden Markov Models (HMMs)
are used to classify metamorphic malware. This technique has
served a benchmark in a variety of other studies~\cite{BLS,Runwal,Shanmugam,Annie}.
Consequently, we use HMMs as the basis for the malware detection schemes
considered in this research.

\subsection{Hidden Markov Models}

A Hidden Markov Model can be viewed as a machine learning
technique, based on a discrete hill climb~\cite{hmm}. 
Applications of HMMs are many and varied, 
ranging from speech recognition to applications in 
computational molecular biology, to artificial intelligence, to malware detection~\cite{hmmintro}. 

As the name suggests, a Hidden Markov Model includes a Markov process that
cannot be directly observed. In an HMM, we have a series of observations
that are related to the ``hidden'' Markov process by a set of discrete probability distributions.

\begin{figure}[htb]
  \begin{center}
    \begin{tikzpicture}[scale=1.0]
    
    \draw[thick,color=blue] (0,0) rectangle (1,1);
    \draw[thick,color=blue] (2.5,0) rectangle (3.5,1);
    \draw[thick,color=blue] (5,0) rectangle (6,1);
    \draw[thick,color=blue] (10,0) rectangle (11,1);

    \draw[thick,color=green] (0.5,4.5) circle (0.575);
    \draw[thick,color=green] (3,4.5) circle (0.575);
    \draw[thick,color=green] (5.5,4.5) circle (0.575);
    \draw[thick,color=green] (10.5,4.5) circle (0.575);
    
    \node at (0.5,0.5){$\O_0$};
    \node at (3,0.5){$\O_1$};
    \node at (5.5,0.5){$\O_2$};
    \node at (8,0.5){$\cdots$};
    \node at (10.5,0.5){$\O_{T-1}$};

    \node at (0.5,4.5){$X_0$};
    \node at (3,4.5){$X_1$};
    \node at (5.5,4.5){$X_2$};
    \node at (8,4.5){$\cdots$};
    \node at (10.5,4.5){$X_{T-1}$};
       
    \node at (1.7,4.8){$A$};
    \node at (4.2,4.8){$A$};
    \node at (6.7,4.8){$A$};
    \node at (9.2,4.8){$A$};
    
    \node at (0.2,2.1){$B$};
    \node at (2.7,2.1){$B$};
    \node at (5.2,2.1){$B$};
    \node at (10.2,2.1){$B$};
    
     \draw[thick,color=black,->] (1.075,4.5) -- (2.425,4.5);
     \draw[thick,color=black,->] (3.575,4.5) -- (4.925,4.5);
     \draw[thick,color=black,->] (6.075,4.5) -- (7.425,4.5);
     \draw[thick,color=black,->] (8.575,4.5) -- (9.925,4.5);

     \draw[thick,color=black,->] (0.5,3.925) -- (0.5,1);
     \draw[thick,color=black,->] (3.0,3.925) -- (3.0,1);
     \draw[thick,color=black,->] (5.5,3.925) -- (5.5,1);
     \draw[thick,color=black,->] (10.5,3.925) -- (10.5,1);

    \draw[thick,dashed,color=red] (-0.3,3) -- (11.2,3);
   
    \end{tikzpicture}
  \end{center}
  \caption{Generic Hidden Markov Model\label{fig:hmm}}
\end{figure}
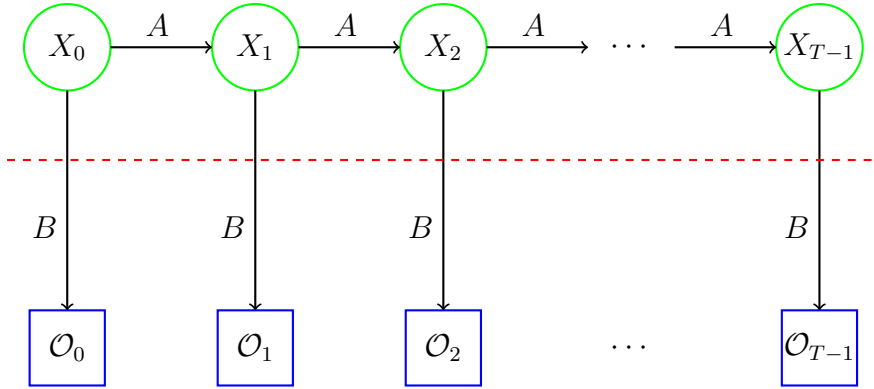

We use the following notation for an HMM~\cite{hmm}:
$$
  \advance\arraycolsep by -2.5pt
  \begin{array}{ccl}
    T & = & \mbox{length of the observation sequence}\\
    N & = & \mbox{number of states in the model}\\
    M & = & \mbox{number of observation symbols}\\
    Q & = & \{q_0,q_1,\ldots,q_{N-1}\} = 
            \mbox{distinct states of the Markov process}\\
    V & = & \{0,1,\ldots,M-1\} = \mbox{set of possible observations}\\
    A & = & \mbox{state transition probabilities}\\
    B & = & \mbox{observation probability matrix}\\
    \pi & = & \mbox{initial state distribution}\\
    \O & = & (\O_0,\O_1,\ldots,\O_{T-1}) = \mbox{observation sequence}.
  \end{array}
$$
A generic Hidden Markov Model is illustrated in Figure~\ref{fig:hmm}.


A Hidden Markov Model is defined by the matrices~$\pi$, $A$ and~$B$,
and hence we denote an HMM as~$\lambda = (A, B, \pi)$.
For simplicity, we often refer to~$\lambda$ simply as a ``model''. 

The practical utility of HMMs derives from the fact that there are
efficient algorithms to solve each of the following three problems~\cite{hmm}.
\begin{description}
\item[Problem 1:] Given a model~$\lambda = (A, B, \pi)$ and an observation 
sequence~$\O$, determine~$P(\O\,|\,\lambda)$. 
That is, we can score a given observation sequence against 
a given model---the better the score, the more closely the observation sequence 
matches the observations used to train the model.
\item[Problem 2:] Given a model~$\lambda = (A, B, \pi)$ and an observation 
sequence~$\O$, determine the optimal state sequence~$X$. 
That is, we can uncover the ``best'' hidden state sequence.
Here, ``best'' is in the sense of maximizing the expected number
of correct states~$X_i$. This is in contrast to a dynamic program, 
which yields the~$X_i$ corresponding to the highest scoring path.
\item[Problem 3:] Given an observation sequence~$\O$ and parameters~$N$ 
and~$M$, determine the model~$\lambda = (A,B,\pi)$ 
such that~$P(\O\,|\,\lambda)$ is maximized. 
That is, we can train a model to fit a given observation sequence~$\O$. 
\end{description}

In this research, we use the solution to Problem~3 to train a model based on observation
sequences extracted from a given malware family. Then we use the solution to Problem~1 to
score observation sequences extracted from malware files as well as sequences extracted from
benign files. We use the resulting scores to measure the success of each technique.

Details on the HMM algorithms are beyond the scope
of this paper. For a thorough discussion of the solutions to HMM Problems~1 through~3,
see~\cite{hmm}; for additional information see~\cite{hmmintro} or the
classic introduction~\cite{LRR}. For the application of HMMs to malware
detection, see, for example,~\cite{wong}.

\subsection{Related Work\label{sect:rw}}

Here we discuss some relevant examples of previous work. We group the previous
work based on whether it relies on static analysis or dynamic analysis, and
we discuss techniques that employ a hybrid approach.

\subsubsection{Static Analysis}

Static analysis of software is performed without actually 
executing the program~\cite{egelesurvey}. Examples of the
information we can obtain from static analysis include
opcode sequences (extracted by disassembling the binary file), 
control flow graphs, and so on. 
Such feature sets can be used individually or in combination 
for malware detection.

In~\cite{mihai}, the authors presented a malware detection technique 
that relies on static analysis and is based on control flow graphs.
Their approach focuses on detecting obfuscation patterns in malware 
and they are able to achieve good accuracy. 

Machine learning techniques have been applied to malware detection 
in the context of static detection. In~\cite{wong}, Hidden Markov Models are
used to effectively classify metamorphic malware, based on extracted
opcode sequences. A similar analysis involving Profile Hidden Markov Models
is considered in~\cite{Attaluri}, while Principal Component Analysis is
used in~\cite{JAS} and~\cite{DPS}, and Support Vector Machines
are used for malware detection in~\cite{Tanuvir}.
The paper~\cite{chinmayee} employs clustering,
based on features derived from static analysis, 
for malware classification. 

In~\cite{pdeshpande}, function call graph analysis is used for malware detection,
while~\cite{Shanmugam} analyzes an opcode-based similarity measure
that relies on simple substitution cryptanalysis techniques. 
API call sequences and opcode sequences are both used 
in~\cite{shankarpani}  to determine whether a segment of code 
has similarity to some particular malware.

The papers~\cite{BLS,LAS} analyze file structure based on entropy variations.
The work in these paper was inspired by the entropy-based score in~\cite{sorokin}.

\subsubsection{Dynamic Analysis}

Dynamic analysis requires that we execute the program, 
often in a virtual environment~\cite{egelesurvey}. 
Examples of information that can be obtained by dynamic analysis include 
API calls, system calls, instruction traces, registry changes, memory writes, and so on. 

In~\cite{kolbitsch}, the authors build fine-grained models 
that are designed to capture the behavior of malware based on system calls. 
The resulting behavior models are represented in the form of graphs, where the
vertices denote system calls and the edges denote dependency between the calls.  

The paper~\cite{ahmed} presents a run-time monitoring tool that extracts statistical 
features based on spatio-temporal information in API call logs. 
The spatial information consists of the arguments and return values of the 
API calls, while the temporal information is the sequencing of the API calls. 
This information is used to build formal models that are fed into standard 
machine learning algorithms, for use in malware detection.

In~\cite{mojtaba1}, a set of program API calls is extracted and combined with 
control flow graphs to obtain a so-called API-CFG model. 
In a slightly modified version~\cite{hdm2},
$n$-gram methods are applied to the API calls. 


Some recent work focuses on kernel execution traces as a means of
developing a malware behavior 
monitor~\cite{rhee}. In~\cite{park}, the authors present an effective method for 
malware classification using graph algorithms, which relies on dynamically 
extracted information. The related work~\cite{youghee}
constructs a ``kernel object behavioral graph'' and graph isomorphism 
techniques are used for scoring.

API sequences are again used for malware detection in~\cite{imds}. 
Also in the paper~\cite{yong}, malware is analyzed based
on frequency analysis of API call sequences.

In~\cite{ratan}, dynamic instruction sequences are logged and converted into abstract 
assembly blocks. Data mining algorithms are used to build a classification model 
that relies on feature vectors extracted from this data.

The authors of~\cite{Anderson} propose a malware detection technique that uses instruction trace 
logs of executables, where this information is collected dynamically. 
These traces are then analyzed as graphs, where 
the instructions are the nodes, and statistics from instruction traces are used to calculate transition
probabilities. Support Vector Machines are used to determine the actual classification. 

\subsubsection{Hybrid Approaches}

Hybrid techniques combine aspects of both static and dynamic analysis. 
In this section, we discuss two recent examples of work of this type.

In~\cite{young}, the authors propose a framework for classification 
of malware using both static and dynamic analysis. 
They define features of malware using an approach that they call
Malware DNA (Mal-DNA). The heart of this technique is 
a debugging-based behavior monitor and analyzer that 
extracts dynamic characteristics.


In the paper~\cite{mojtabahdm}, the authors develop and analyze a tool that they call 
HDM Analyser. This tool uses both static analysis and dynamic analysis
in the training phase, but performs only static analysis in the testing phase. 
The goal is to take advantage of the supposedly superior fidelity
of dynamic analysis in the training phase, while maintaining the efficiency
advantage of static detection in the scoring phase. 
For comparison, it is shown that HDM Analyser has better overall accuracy and time
complexity than the static or dynamic analysis methods in~\cite{mojtabahdm}. 
The dynamic analysis in~\cite{mojtabahdm} is based on  extracted API call sequences.

Next, we discuss ROC analysis. We use the area under the ROC curve as one of our
measures of success for the experiments reported in Section~\ref{sect:experiments}.

\subsection{ROC Analysis\label{sect:ROC}}

A Receiver Operating Characteristic (ROC) curve is 
obtained by plotting the false positive rate against the true positive rate
as the threshold varies through the range of data values. An 
Area Under the ROC Curve (AUC-ROC) of~1.0 implies ideal detection, that is,
there exists a threshold for which no false positives or false negatives occur.
The AUC-ROC can be interpreted as the probability that a randomly selected
positive instance scores higher than a randomly selected negative instance~\cite{bradley,rocintro}.
Therefore, an AUC-ROC of~0.5 means that the binary classifier is no better than flipping a coin.
Also, an AUC-ROC that is less than~0.5 implies that we can obtain a classifier with an AUC-ROC
greater than~0.5 by simply reversing the classification criteria. 

\begin{figure}[htbp]
  \begin{center}
    \begin{tikzpicture}[scale=0.8]
        
    \draw[thick,color=red,fill=red] (0.5,3) circle (0.08);
    \draw[thick,color=red,fill=red] (1.0,4.25) circle (0.08);
    \draw[thick,color=red,fill=red] (1.5,2.0) circle (0.08); 
    \draw[thick,color=red,fill=red] (2.0,2.75) circle (0.08);
    \draw[thick,color=red,fill=red] (2.5,2.65) circle (0.08);
    \draw[thick,color=red,fill=red] (3.0,2.7) circle (0.08);
    \draw[thick,color=red,fill=red] (3.5,1.0) circle (0.08); 
    \draw[thick,color=red,fill=red] (4.0,2.5) circle (0.08);
    \draw[thick,color=red,fill=red] (4.5,2.1) circle (0.08); 
    \draw[thick,color=red,fill=red] (5.0,2.75) circle (0.08);

    \draw[thick,color=blue] (0.5,1.75) rectangle (0.65,1.9);
    \draw[thick,color=blue] (1.0,1.25) rectangle (1.15,1.4);
    \draw[thick,color=blue] (1.5,1.5) rectangle (1.65,1.65);
    \draw[thick,color=blue] (2.0,4.1) rectangle (2.15,4.25); 
    \draw[thick,color=blue] (2.5,0.5) rectangle (2.65,0.65);
    \draw[thick,color=blue] (3.0,1.5) rectangle (3.15,1.65);
    \draw[thick,color=blue] (3.5,1.85) rectangle (3.65,2.0);
    \draw[thick,color=blue] (4.0,3.5) rectangle (4.15,3.65); 
    \draw[thick,color=blue] (4.5,0.95) rectangle (4.65,1.1);
    \draw[thick,color=blue] (5.0,1.45) rectangle (5.15,1.6);

    \draw[ultra thick,color=yellow,dashed] (0,2.25) -- (5.5,2.25); 
    
    \draw[thick,color=black,->] (0,0) -- (6,0); 
    \draw[thick,color=black,->] (0,0) -- (0,5); 


    \draw[ultra thick,color=red] (8.0,0.0) -- (8.0,0.5) -- (9.0,0.5) -- (9.0,4.5) -- (12.5,4.5) -- (12.5,5.0) -- (13.0,5.0);
    \draw[fill=red,opacity=0.2] plot coordinates {(8.0,0.0) (8.0,0.5) (9.0,0.5) (9.0,4.5) (12.5,4.5) (12.5,5.0) (13.0,5.0) (13.0,0.0)};

    \draw[thick,color=black,fill=black] (9,3.5) circle (0.08);
    
    \draw[thick,color=black] (8,0) -- (13,0); 
    \node[rotate=0] at (10.5,-0.4){$1 - \mbox{specificity}$};
    \draw[thick,color=black] (8,0) -- (8,5); 
    \node[rotate=90] at (7.6,2.5){$\mbox{sensitivity}$};
    \draw[thick,color=black] (8,5) -- (13,5);
    \draw[thick,color=black] (13,5) -- (13,0);

    \draw[thick,color=black] (8,0) -- (7.85,0);
    \node[rotate=0] at (7.65,0){$0$};
    \draw[thick,color=black] (8,0) -- (8,-0.15);
    \node[rotate=0] at (8.05,-0.4){$0$};
    \draw[thick,color=black] (8,5) -- (7.85,5);
    \node[rotate=0] at (7.65,5){$1$};
    \draw[thick,color=black] (13,0) -- (13,-0.15);
    \node[rotate=0] at (13.05,-0.4){$1$};


    \end{tikzpicture}
  \end{center}
  \caption{Scatterplot and ROC Curve\label{fig:dataROCpt}}
\end{figure}
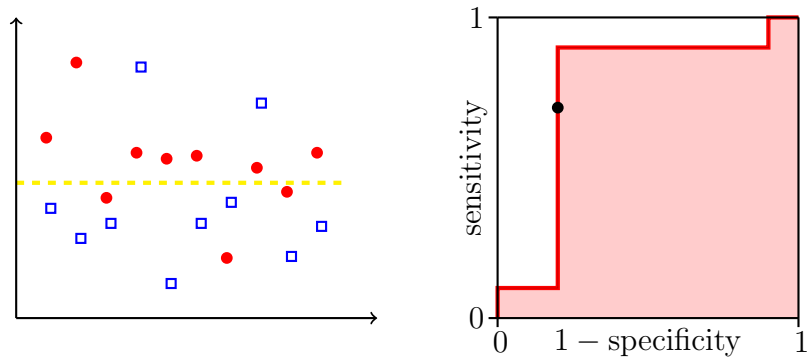

An examples of a scatterplot and the corresponding ROC curve 
is given in Figure~\ref{fig:dataROCpt}. The red circles in the scatterplot
represent positive instances, while the blue squares represent negative
instances. In the context of malware classification, the red circles are
scores for malware files, while the blue squares represent scores for benign files.
Furthermore, we assume that higher scores are ``better'', that is, for this 
particular score, positive instances are supposed to
score higher than negative instances.

For a given experiment, the true positive rate is also known as the sensitivity, while
the true negative rate is referred to as the specificity. Then the false positive rate is
given by~$1-\mbox{specificity}$.

Note that if we place the threshold below the lowest point in the scatterplot 
in Figure~\ref{fig:dataROCpt}, then
$$
  \mbox{sensitivity} = 1 \mbox{\ \ and\ \ } 1 - \mbox{specificity} = 1
$$
On the other hand, if we place the threshold above the highest point, then
$$
  \mbox{sensitivity} = 0 \mbox{\ \ and\ \ } 1 - \mbox{specificity} = 0
$$
Consequently, an ROC curve must always include the points~$(0,0)$
and~$(1,1)$. The intermediate points on the ROC curve are determined as the threshold
passes through the range of values. For example, if we place the threshold at the 
yellow dashed line in the scatterplot in Figure~\ref{fig:dataROCpt}, the true positive rate 
(i.e., sensitivity) is~0.7, since~7 of the~10 positive instances are
classified correctly, while the false positive rate (i.e., $1 - \mbox{specificity}$) is~0.2,
since~2 of the~10 negative cases lie on the wrong side of the threshold. This implies
that the point~$(0.2,0.7)$ lies on the ROC curve. The point~$(0.2,0.7)$ 
is illustrated by the black circle on the ROC graph in Figure~\ref{fig:dataROCpt}.
The shaded region in Figure~\ref{fig:dataROCpt} represents the AUC. In this example, 
we find that the AUC-ROC is~0.75.


\subsection{PR Analysis\label{sect:PR}}

Precision Recall (PR) curves offer an alternative to ROC analysis for scatterplot data~\cite{PR_curves}. 
There are many connections between PR curves and ROC 
curves,\footnote{For example, if one curve dominates another in ROC
space, it also dominates in PR space, and vice versa.}
but in certain cases, PR curves can be more informative. In particular, when the nomatch
set is large relative to the match set, PR curves may be preferred. 

We define~$\rec$ to be the fraction of the match cases that are classified correctly,
and~$\prec$ is the fraction of elements classified as positive that actually
belong to the match set. More precisely, 
$$
  \rec = \frac{\TP}{\TP + \FN} \mbox{\ \ and\ \ } \prec = \frac{\TP}{\TP + \FP} 
$$
where~$\TP$ is the number of true positives, $\FP$ is the number of false positives, 
and~$\FN$ is the number of false negatives. 
Note that~$\rec$ is the true positive rate, which we referred to
as sensitivity in our discussion of ROC curves.
However, $\prec$ is not the same as the false positive rate 
which is used to compute ROC curves.
Also note that~$\TN$ does not appear in the formula for~$\rec$
or~$\prec$, and hence 
true negatives play no (direct) role in computing the PR curve. Again, this may
be useful if we want to focus our attention on the positive set, particularly
when we have a relatively large negative set.
As with ROC analysis, we can use the Area Under the PR Curve (AUC-PR)
as a measure of the success of a classifier.

\begin{figure}[htbp]
  \begin{center}
    \begin{tikzpicture}[scale=0.8]
        
    \draw[thick,color=red,fill=red] (0.5,3) circle (0.08);
    \draw[thick,color=red,fill=red] (1.0,4.25) circle (0.08);
    \draw[thick,color=red,fill=red] (1.5,2.0) circle (0.08); 
    \draw[thick,color=red,fill=red] (2.0,2.75) circle (0.08);
    \draw[thick,color=red,fill=red] (2.5,2.65) circle (0.08);
    \draw[thick,color=red,fill=red] (3.0,2.7) circle (0.08);
    \draw[thick,color=red,fill=red] (3.5,1.0) circle (0.08); 
    \draw[thick,color=red,fill=red] (4.0,2.5) circle (0.08);
    \draw[thick,color=red,fill=red] (4.5,2.1) circle (0.08); 
    \draw[thick,color=red,fill=red] (5.0,2.75) circle (0.08);

    \draw[thick,color=blue] (0.5,1.75) rectangle (0.65,1.9);
    \draw[thick,color=blue] (1.0,1.25) rectangle (1.15,1.4);
    \draw[thick,color=blue] (1.5,1.5) rectangle (1.65,1.65);
    \draw[thick,color=blue] (2.0,4.1) rectangle (2.15,4.25); 
    \draw[thick,color=blue] (2.5,0.5) rectangle (2.65,0.65);
    \draw[thick,color=blue] (3.0,1.5) rectangle (3.15,1.65);
    \draw[thick,color=blue] (3.5,1.85) rectangle (3.65,2.0);
    \draw[thick,color=blue] (4.0,3.5) rectangle (4.15,3.65); 
    \draw[thick,color=blue] (4.5,0.95) rectangle (4.65,1.1);
    \draw[thick,color=blue] (5.0,1.45) rectangle (5.15,1.6);

    \draw[ultra thick,color=yellow,dashed] (0,2.25) -- (5.5,2.25); 
    
    \draw[thick,color=black,->] (0,0) -- (6,0); 
    \draw[thick,color=black,->] (0,0) -- (0,5); 


    
     \draw[ultra thick,color=red] (13.000000,2.500000) -- (13.000000,2.631579) -- (13.000000,2.777778) 
     	-- (12.500000,2.647059) -- (12.500000,2.812500) -- (12.500000,3.000000) -- (12.500000,3.000000)
	-- (12.500000,3.461538) -- (12.500000,3.750000) -- (12.500000,4.090909) -- (12.000000,4.000000) 
	-- (11.500000,3.888889) -- (11.000000,3.750000) -- (10.500000,3.571429) -- (10.000000,3.333333) 
	-- (10.000000,3.333333) -- (9.000000,2.500000) -- (8.500000,1.666667) -- (8.500000,2.500000) 
	-- (8.500000,5.000000) -- (8.000000,5.000000);
    \draw[fill=red,opacity=0.2] plot coordinates {(8.0,0.0) (8.0,5.0) (8.500000,5.000000) (8.500000,1.666667)
   	 (9.000000,2.500000) (10.000000,3.333333) (10.500000,3.571429) (11.000000,3.750000) (11.500000,3.888889)
	 (12.000000,4.000000) (12.500000,4.090909) (12.500000,2.647059) (13.000000,2.777778) (13.000000,0.0000)};

    \draw[thick,color=black,fill=black] (11.5,3.9) circle (0.08);
    
    \draw[thick,color=black] (8,0) -- (13,0); 
    \node[rotate=0] at (10.5,-0.4){$\rec$};
    \draw[thick,color=black] (8,0) -- (8,5); 
    \node[rotate=90] at (7.6,2.5){$\prec$};
    \draw[thick,color=black] (8,5) -- (13,5);
    \draw[thick,color=black] (13,5) -- (13,0);

    \draw[thick,color=black] (8,0) -- (7.85,0);
    \node[rotate=0] at (7.65,0){$0$};
    \draw[thick,color=black] (8,0) -- (8,-0.15);
    \node[rotate=0] at (8.05,-0.4){$0$};
    \draw[thick,color=black] (8,5) -- (7.85,5);
    \node[rotate=0] at (7.65,5){$1$};
    \draw[thick,color=black] (13,0) -- (13,-0.15);
    \node[rotate=0] at (13.05,-0.4){$1$};


    \end{tikzpicture}
  \end{center}
  \caption{Scatterplot and PR Curve\label{fig:dataPRpt}}
\end{figure}
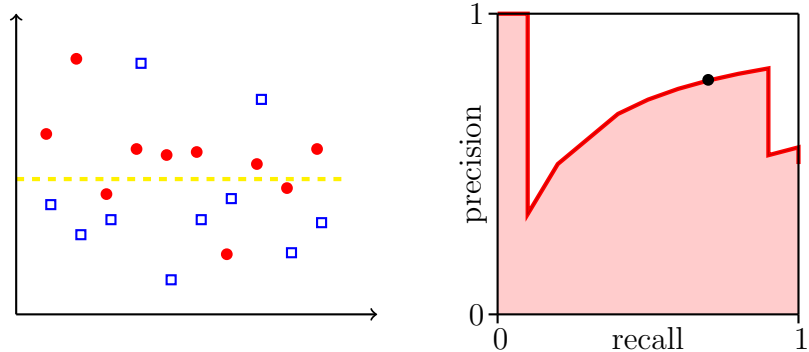

To generate the PR curve, we plot the~$(\rec,\prec)$ pairs as the 
threshold varies through the range of values in a given scatterplot.
To illustrate the process, we consider the same data as in the
ROC curve example in Section~\ref{sect:ROC}.
This data and the corresponding PR curve is given
here in Figure~\ref{fig:dataPRpt}.

For the threshold that appears in the scatterplot in Figure~\ref{fig:dataPRpt}, 
we have~$\TP=7$, $\FP=2$, and~$\FN=3$, and hence
$$
  \rec = \frac{7}{7 + 3} = 0.7 \mbox{\ \ and\ \ } \prec = \frac{7}{7 + 2} \approx 0.78
$$
This point is plotted on the right-hand side of Figure~\ref{fig:dataPRpt}
and the entire PR curve is given. In this example, the AUC-PR is about~0.69.

\section{Experiments\label{sect:implementation}}

In this section, we first discuss the tools that we use to extract features
from code---both statically
and dynamically. Then we give an
overview of the malware dataset used in this project. 
Finally, we elaborate on the experiments conducted. 
In Section~\ref{sect:experiments}, 
we provide the results of our experiments.

\subsection{Tools for Dynamic and Static Analysis\label{sect:tools}}


IDA Pro is a disassembler that generates highly accurate assembly code from an
executable. It can also be used as a debugger. IDA is a powerful tool
that supports scripting, function tracing, instruction tracing, instruction logging, etc.
In this research, we use IDA Pro for static analysis, specifically, 
to generate {\tt .asm} files from {\tt .exe} files,
from which opcodes and windows API calls can be extracted. 
We also use IDA Pro for dynamic analysis, specifically, to collect instruction traces
from executables. 

In addition to IDA Pro, for dynamic analysis we use 
the Buster Sandbox Analyzer (BSA).
BSA is a dynamic analysis tool that has been designed to 
determine if a process exhibits potentially malicious behavior. In addition to analyzing the 
behavior of a process, BSA keeps track of actions taken by a monitored program,
such as registry changes, file-system changes, and port changes~\cite{bsa}. 
The tool runs inside a sandbox which protects the system from infection 
while executing malware. The sandbox used by BSA is known
as Sandboxie~\cite{sandboxie}. 

For dynamic analysis, we also experimented with Ether~\cite{ether}.
Ether is an open source tool that resides completely outside of the target OS,
which makes it difficult for a program to detect
that emulation is occurring. This is potentially useful for malware analysis,
since viruses can, in principle, detect a debugger or 
a virtual environment during execution. However, for the
datasets considered in this paper,
we found no significant differences in the API call sequences 
generated by BSA and Ether. Since BSA is more user-friendly, 
in this research, we exclusively used BSA to generate our dynamic API call sequences. 


\subsection{Datasets}

The following seven malware families were used as datasets 
in this research~\cite{NRC}. 

\begin{description}
\item[Harebot] is a backdoor that provides remote access to the infected system. 
Because of its many features, it is also considered to be a rootkit~\cite{harebot}. 
\item[Security Shield] is a Trojan that, like Winwebsec, claims to be anti-virus software.
Security Shield reports fake virus detection messages and attempts to
coerce the users into purchasing software~\cite{securityshield}.
\item[Smart HDD] reports various problems with the hard drive and tries to
convince the user to purchase a product to fix these ``errors''. 
Smart HDD is named after S.M.A.R.T., which is 
a legitimate tool that monitors hard disk drives (HDDs)~\cite{smarthdd}.
\item[Winwebsec] pretends to be anti-virus software. 
An infected system displays fake messages claiming malicious activity 
and attempts to convince the user to pay money for software to clean the 
supposedly infected system~\cite{winwebsec}.
\item[Zbot] also known as Zeus, is a Trojan horse that compromises a system 
by downloading configuration files or updates. 
Zbot is a stealth virus that hides in the file system~\cite{zbot}. 
The virus eventually vanishes from the processes list and, consequently, 
we could only trace its execution for about~5 to~10 minutes.
\item[ZeroAccess] is a Trojan horse that makes use of an advanced rootkit to hide itself. 
ZeroAccess is capable of creating a new hidden file system, it can
create a backdoor on the compromised 
system, and it can download additional malware~\cite{zeroaccess}.
\end{description}
Table~\ref{malware-dataset} gives the number of files used 
from each malware family and the benign dataset. 
\begin{table}[htb]
\caption{Datasets\label{malware-dataset}}
\begin{center}
\begin{tabular}{l|c}
\hline\hline
\multirow{2}{*}{Family} & \multicolumn{1}{c}{Number} \\
 & \multicolumn{1}{c}{of Files} \\ \hline
Harebot & \z45  \\
Security Shield & \z50\\
Smart HDD & \z50\\
Winwebsec & 200  \\
Zbot & 200  \\
ZeroAccess & 200 \\
\hline
benign & \z40 \\ \hline\hline
\end{tabular}
\end{center}
\end{table}
For our benign dataset, we use the set of Windows System 32 files
listed in Table~\ref{tab:benign}. 

\begin{table}[htbp]
\caption{Benign Dataset\label{tab:benign}}
\begin{center}
{\tt 
\begin{tabular}{llll}
\hline
notepad & alg & calc & cipher \\
cleanmgr & cmd & cmdl32 & driverquery \\
drwtsn32 & dvdplay & eventcreate & eventtriggers \\
eventvwr & narrator & freecell & grpconv \\
mshearts & mspaint & netstat & nslookup \\
osk & packager & regedit & sndrec32 \\
sndvol32 & sol & sort & spider \\
syncapp & ipconfig & taskmgr & telnet \\
verifier & winchat & charmap & clipbrd \\
ctfmon & wscript & mplay32 & winhlp32 \\
\hline
\end{tabular}
}
\end{center}
\end{table}

\subsection{Data Collection}

For training and scoring, we use opcode sequences and API calls.
For both opcode and API call sequences, we extract the data using 
both a static and a dynamic approach, giving us four
observation sequences for each program under consideration. 
As noted in Section~\ref{sect:rw}, opcode sequences and
API call traces have been used in many research studies on 
malware detection.

We use IDA Pro to disassemble files and we extract the static opcode sequences
from the resulting disassembly. 
For example, suppose we disassemble an {\tt exe} file
and obtain the disassembly in Table~\ref{tab:disass}.
\begin{table}
\caption{Example Disassembly\label{tab:disass}}
\begin{center}
\def\u{\underline{\phantom{M}}}
\tt
\begin{tabular}{lcl}\hline
.text:00401017 &         & call  sub\u401098\\
.text:0040101C &        & push  8\\
.text:0040101E &        & lea   ecx, [esp+24h+var\u14]\\
.text:00401022 &        & push  offset xyz\\
.text:00401027 &        & push  ecx\\
.text:00401028 &        & call  sub\u401060\\
.text:0040102D &       & add   esp, 18h\\
.text:00401030 &        & test  eax, eax\\
.text:00401032 &        & jz    short loc\u401045\\
\hline
\end{tabular}
\end{center}
\end{table}
The static opcode sequence corresponding to this disassembly is
\begin{center}
{\tt call}, {\tt push}, {\tt lea}, {\tt push}, {\tt push}, {\tt call}, {\tt add}, {\tt test}, {\tt jz}
\end{center}
We discard all operands, labels, directives, etc., and only retain 
the mnemonic opcodes.

For dynamic opcode sequences, we execute the program in IDA Pro 
using the ``tracing'' feature.
From the resulting program trace, we extract mnemonic opcodes. 
Note that the static opcode sequence corresponds 
to the overall program structure, while the dynamic opcode sequence corresponds
to the actual execution path taken when the program was traced.

Microsoft Windows provides a variety of API (Application Programming Interface)
calls that facilitate requests for services 
from the operating system~\cite{ahmed}. 
Each API call has a distinct name, a set of arguments, and a return value. 
We only collect API call names, discarding the arguments and return value. 
An example of a sequence of API calls is given by
\begin{center}
{\tt OpenMutex}, 
{\tt CreateFile}, 
{\tt OpenProcessToken}, 
{\tt AdjustTokenPrivileges}, 
{\tt SetNamedSecurityInfo}, 
{\tt LoadLibrary}, 
{\tt CreateFile}, 
{\tt GetComputerName}, 
{\tt QueryProcessInformation}, 
{\tt VirtualAllocEx}, 
{\tt DeleteFile}
\end{center}

As with opcodes, API calls can be extracted from executables 
statically or dynamically. 
Our static API call sequences are obtained from IDA Pro disassembly.
As mentioned in Section~\ref{sect:tools}, 
we use Buster Sandbox Analyser (BSA) to dynamically extract API calls. 
BSA allows us to execute a program for a fixed amount of time, and 
it logs all API calls that occur within this execution window. 
From these logged API calls, we form a dynamic API call sequence for each 
executable.

\subsection{Training and Scoring\label{sect:Trainingcases}}

For our experiments, four cases are considered. In the first, we use the static 
observation sequences for both training and scoring. In the second case, 
we use the dynamically extracted data for both training and scoring.
The third and fourth cases are hybrid situations. Specifically, in the
third case, we use the dynamic data for training, 
but the static data for scoring. In the fourth case, we use static training data,
but dynamic data for scoring. We denote these four cases
as static/static, dynamic/dynamic, static/dynamic, and dynamic/static, 
respectively.

Our static/static and dynamic/dynamic cases can be viewed as representative of
typical approaches used in static and dynamic detection. The 
dynamic/static case is analogous to the approach used in many
hybrid schemes. This approach seems to offer the prospect
of the best of both worlds. That is, we can have a more accurate model
due to the use dynamic training data, and yet scoring remains efficient,
thanks to the use of static scoring data.
Since the training phase is 
essentially one-time work, it is acceptable to spend significant 
time and effort in training. And the scoring phase can be no better
than the model generated in the training phase.

On the other hand, the static/dynamic seems to offer no clear advantage.
For completeness, we include this case in our opcode experiments. 

We conducted a separate experiment for each of the malware 
datasets listed in Table~\ref{malware-dataset}, 
for each of the various combinations of static and dynamic data
mentioned above.
For every experiment, we use five-fold cross validation. That is, the 
malware dataset is partitioned into five equal subsets, say, $S_1$, $S_2$,
$S_3$, $S_4$, and~$S_5$. Then 
subsets~$S_1$, $S_2$, $S_3$, and~$S_4$ are used to train an HMM, 
and the resulting model is used to score the malware in~$S_5$,
and to score the files in the benign set.
The process is repeated five times, with a different subset~$S_i$
reserved for testing in each of the five ``folds''.  Cross validation
serves to smooth out any bias in the partitioning of the data, while
also maximizing the number of scores obtained from the 
available data. 

The scores from a given experiment
are used to form a scatterplot, from which an ROC curve is
generated. The area under the ROC curve serving as our
measure of success, as discussed in Section~\ref{sect:ROC}.

\section{Results\label{sect:experiments}}

In this section, we present our experimental results. 
We performed experiments with API call sequences 
and separate experiments using 
opcode sequences.
All experiments were conducted as discussed in Section~\ref{sect:Trainingcases}.
That is, different combinations of static and dynamic data
were used for training and scoring. Also, 
each experiment is based on training and scoring with  HMMs, 
using five-fold cross validation.

As discussed in Section~\ref{sect:ROC},
the effectiveness of each experiment is quantified
using the area under the ROC curve (AUC). In this section, we
present AUC results, omitting the scatterplots and ROC curves.
For additional details and results, see~\cite{Anusha}.

\subsection{API Call Sequences}

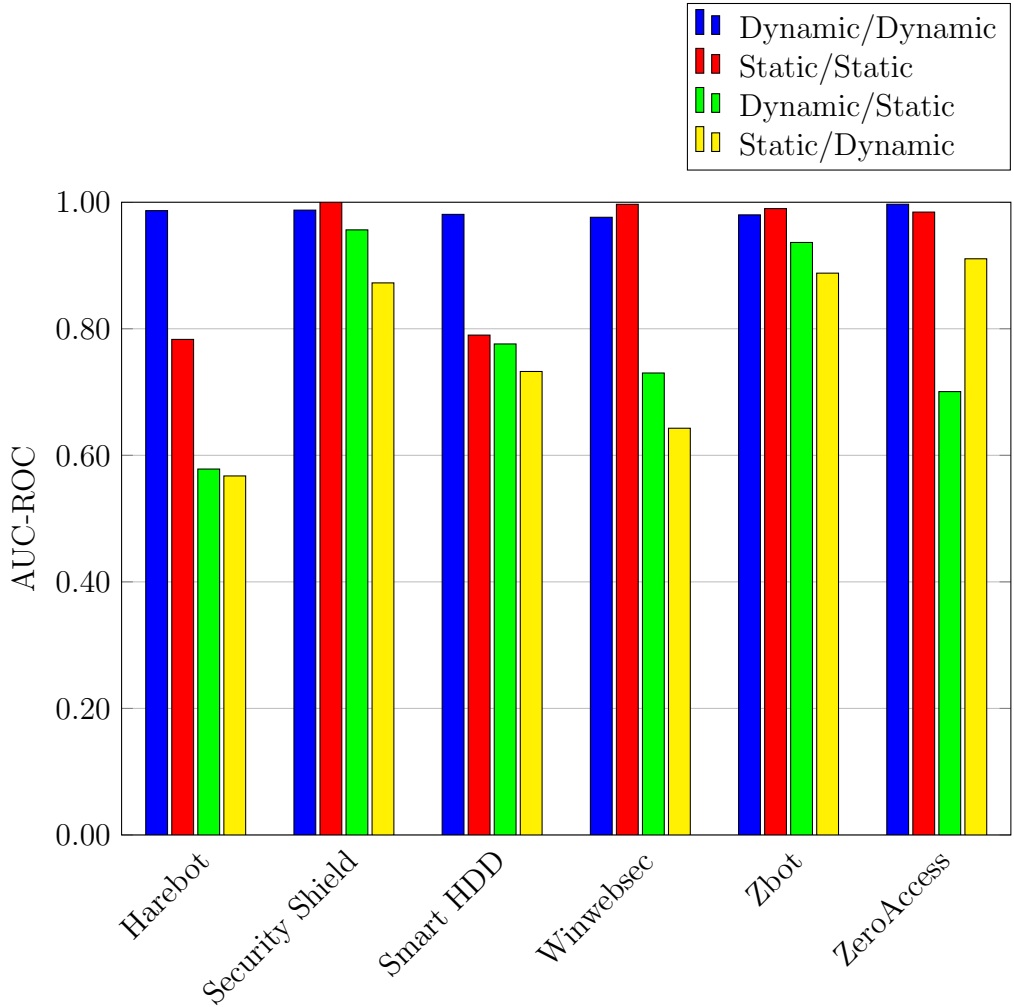
\begin{figure}[htbp]
\begin{center}
\begin{tikzpicture}
    \begin{axis}[
        width  = 0.8*\textwidth,
        height = 10cm,
        ymin=0.0,ymax=1.0,
        major x tick style = transparent,
        ybar=4*\pgflinewidth,
        bar width=8.25pt,
        ymajorgrids = true,
        ylabel = {AUC-ROC},
        symbolic x coords={Harebot,Security Shield,Smart HDD,Winwebsec,Zbot,ZeroAccess},
	y tick label style={
    	/pgf/number format/.cd,
   	fixed,
   	fixed zerofill,
    	precision=2},
        xtick = data,
        x tick label style={rotate=45,anchor=north east, inner sep=0mm},
        scaled y ticks = false,
        enlarge x limits=0.1,
        ymin=0,
        legend cell align=left,
        legend style={
                at={(1,1.05)},
                anchor=south east,
                column sep=1ex
        }
    ]
        \addplot[fill=blue]
            coordinates {
(Harebot,0.9867)
(Security Shield,0.9875)
(Smart HDD,0.9808)
(Winwebsec,0.9762)
(Zbot,0.9800)
(ZeroAccess,0.9968)
};
        \addplot[fill=red]
            coordinates {
(Harebot,0.7832)
(Security Shield,1.0000)
(Smart HDD,0.7900)
(Winwebsec,0.9967)
(Zbot,0.9899)
(ZeroAccess,0.9844)
};
        \addplot[fill=green]
            coordinates {
(Harebot,0.5783)
(Security Shield,0.9563)
(Smart HDD,0.7760)
(Winwebsec,0.7301)
(Zbot,0.9364)
(ZeroAccess,0.7007)
};
        \addplot[fill=yellow]
            coordinates {
(Harebot,0.5674)
(Security Shield,0.8725)
(Smart HDD,0.7325)
(Winwebsec,0.6428)
(Zbot,0.8879)
(ZeroAccess,0.9106)
};
        \legend{Dynamic/Dynamic, Static/Static, Dynamic/Static, Static/Dynamic}
    \end{axis}
\end{tikzpicture}
\end{center}
\vglue-0.3in
\caption{ROC Results for API Call Sequence\label{AUC_bar_API}} 
\end{figure}

We trained HMM models on API call sequences for each of 
the malware families in Table~\ref{malware-dataset}. 
The ROC results are given in Table~\ref{auc-api}, with these same results
plotted in the form of a bar graph in Figure~\ref{AUC_bar_API}.

\begin{table}[htbp]
\caption{AUC-ROC Results for API Call Sequence\label{auc-api}}
\begin{center}
\begin{tabular}{l | cccc}\hline\hline
\multirow{2}{*}{Family} 
& Dynamic/ & Static/ & Dynamic/ & Static/ \\ 
 & Dynamic & Static & Static & Dynamic\\ \hline
Harebot
 & 0.9867
 & 0.7832
 & 0.5783
 & 0.5674
\\
Security Shield
 & 0.9875
 & 1.0000
 & 0.9563
 & 0.8725
\\
Smart HDD
 & 0.9808
 & 0.7900
 & 0.7760
 & 0.7325
\\
Winwebsec
 & 0.9762
 & 0.9967
 & 0.7301
 & 0.6428
\\
Zbot
 & 0.9800
 & 0.9899
 & 0.9364
 & 0.8879
\\
ZeroAccess
 & 0.9968
 & 0.9844
 & 0.7007
 & 0.9106
\\\hline\hline
\end{tabular}
\end{center}
\end{table}

Overall, we see that using dynamic training and testing yields the best results,
while static training and testing is as effective in all cases, except
for Harebot and Smart HDD. Perhaps surprisingly, the hybrid approach of
dynamic training with static scoring
produces worse results than the fully static
case for all families. In fact, the dynamic/static case fares
significantly worse than the static/static
case for all families except Security Shield and Zbot.

\begin{figure}[htbp]
\begin{center}
\begin{tikzpicture}
    \begin{axis}[
        width  = 0.8*\textwidth,
        height = 10cm,
        ymin=0.0,ymax=1.0,
        major x tick style = transparent,
        ybar=4*\pgflinewidth,
        bar width=8.25pt,
        ymajorgrids = true,
        ylabel = {AUC-PR},
        symbolic x coords={Harebot,Security Shield,Smart HDD,Winwebsec,Zbot,ZeroAccess},
	y tick label style={
    	/pgf/number format/.cd,
   	fixed,
   	fixed zerofill,
    	precision=2},
        xtick = data,
        x tick label style={rotate=45,anchor=north east, inner sep=0mm},
        scaled y ticks = false,
        enlarge x limits=0.1,
        ymin=0,
        legend cell align=left,
        legend style={
                at={(1,1.05)},
                anchor=south east,
                column sep=1ex
        }
    ]
        \addplot[fill=blue]
            coordinates {
(Harebot,0.9858)
(Security Shield,0.9884)
(Smart HDD,0.9825)
(Winwebsec,0.9800)
(Zbot,0.9808)
(ZeroAccess,0.9980)
};
        \addplot[fill=red]
            coordinates {
(Harebot,0.8702)
(Security Shield,1.0000)
(Smart HDD,0.8799)
(Winwebsec,0.9967)
(Zbot,0.9931)
(ZeroAccess,0.9879)
};
        \addplot[fill=green]
            coordinates {
(Harebot,0.7111)
(Security Shield,0.9534)
(Smart HDD,0.3768)
(Winwebsec,0.7359)
(Zbot,0.9513)
(ZeroAccess,0.4190)
};
        \addplot[fill=yellow]
            coordinates {
(Harebot,0.4888)
(Security Shield,0.3312)
(Smart HDD,0.4025)
(Winwebsec,0.3947)
(Zbot,0.3260)
(ZeroAccess,0.3472)
};
        \legend{Dynamic/Dynamic, Static/Static, Dynamic/Static, Static/Dynamic}
    \end{axis}
\end{tikzpicture}
\end{center}
\vglue-0.3in
\caption{PR Results for API Call Sequence\label{AUC_PR_bar_API}} 
\end{figure}
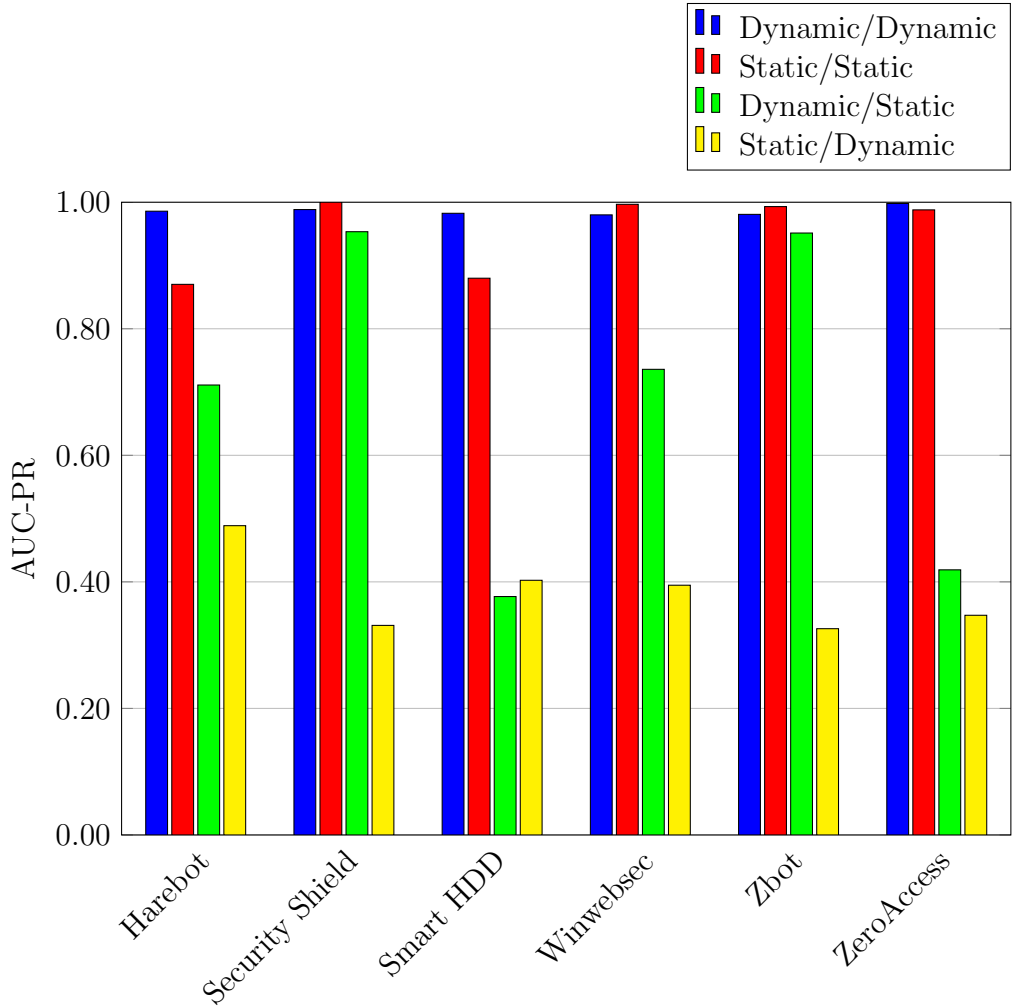

We also computed PR curves for each of 
the malware families in Table~\ref{malware-dataset}. 
The AUC-PR results are given in Table~\ref{auc_pr-api}, with these same results
plotted in the form of a bar graph in Figure~\ref{AUC_PR_bar_API}.

\begin{table}[htbp]
\caption{AUC-PR Results for API Call Sequence\label{auc_pr-api}}
\begin{center}
\begin{tabular}{l | cccc}\hline\hline
\multirow{2}{*}{Family} 
& Dynamic/ & Static/ & Dynamic/ & Static/ \\ 
 & Dynamic & Static & Static & Dynamic\\ \hline
Harebot
 & 0.9858
 & 0.8702
 & 0.7111
 & 0.4888
\\
Security Shield
 & 0.9884
 & 1.0000
 & 0.9534
 & 0.3312
\\
Smart HDD
 & 0.9825
 & 0.8799
 & 0.3768
 & 0.4025
\\
Winwebsec
 & 0.9800
 & 0.9967
 & 0.7359
 & 0.3947
\\
Zbot
 & 0.9808
 & 0.9931
 & 0.9513
 & 0.3260
\\
ZeroAccess
 & 0.9980
 & 0.9879
 & 0.4190
 & 0.3472
\\\hline\hline
\end{tabular}
\end{center}
\end{table}

Next, we provide results for analogous experiments using opcode sequences. 
Then we discuss the significance of these results with respect to static, dynamic,
and hybrid detection strategies.

\subsection{Opcode Sequences}

In this set of experiments, we use opcode sequences for training and scoring.
As in the API sequence case, we consider combinations of static and 
dynamic data for training and scoring. Also as above, we train HMMs and use the resulting 
models for scoring. 

\begin{figure}[htbp]
\begin{center}
\begin{tikzpicture}
    \begin{axis}[
        width  = 0.8*\textwidth,
        height = 8cm,
        major x tick style = transparent,
        ybar=5*\pgflinewidth,
        bar width=14pt,
        ymajorgrids = true,
        ylabel = {Distinct Opcodes},
        symbolic x coords={Harebot,Security Shield,Smart HDD,Winwebsec,Zbot,ZeroAccess},
	y tick label style={
    	/pgf/number format/.cd,
   	fixed,
   	fixed zerofill,
    	precision=0},
        xtick = data,
        x tick label style={rotate=45,anchor=north east, inner sep=0mm},
        scaled y ticks = false,
        enlarge x limits=0.1,
        ymin=0,
        legend cell align=left,
        legend style={
                at={(1,1.05)},
                anchor=south east,
                column sep=1ex
        }
    ]
        \addplot[fill=blue]
            coordinates {
(Harebot,82)
(Security Shield,44)
(Smart HDD,34)
(Winwebsec,49)
(Zbot,42)
(ZeroAccess,52)
};
        \addplot[fill=red]
            coordinates {
(Harebot,207)
(Security Shield,177)
(Smart HDD,81)
(Winwebsec,191)
(Zbot,243)
(ZeroAccess,302)
};
        \legend{Dynamic, Static}
    \end{axis}
\end{tikzpicture}
\end{center}
\vglue-0.3in
\caption{Distinct Opcodes\label{Distinct_Opcodes}} 
\end{figure}
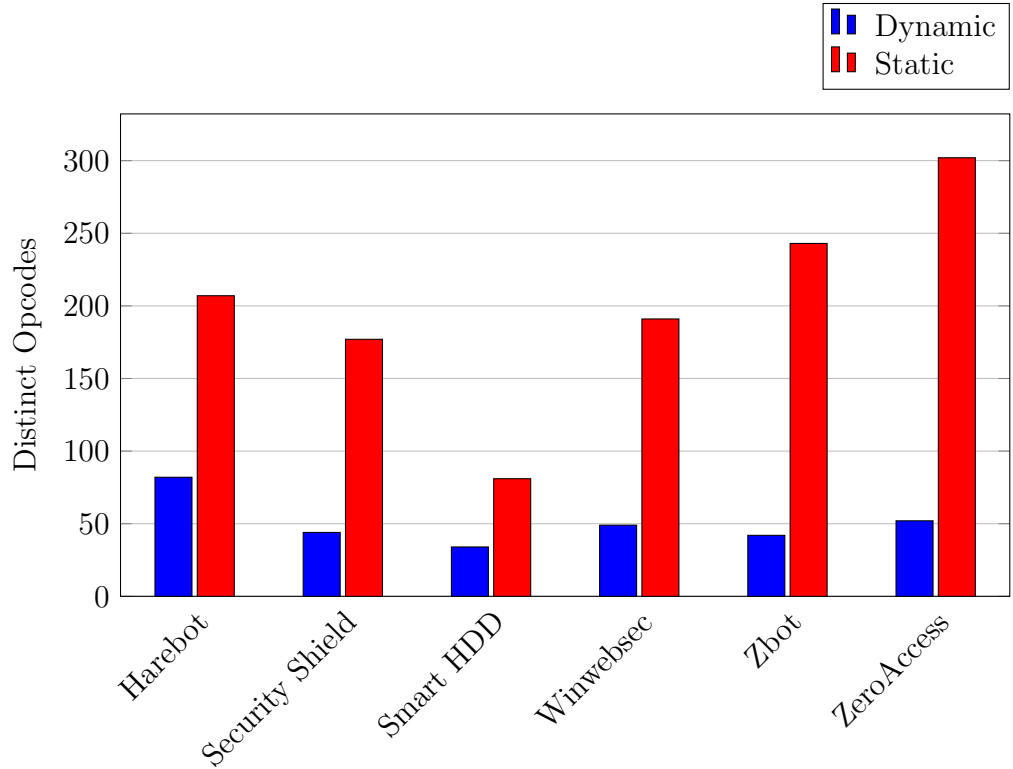

Before presenting our results, we note that the opcode sequences obtained in
the static and dynamic cases differ significantly.
In Figure~\ref{Distinct_Opcodes} we give a bar graph
showing the counts for the number of 
distinct opcodes in the static and dynamic cases. 
From Figure~\ref{Distinct_Opcodes}, we see
that scoring in the dynamic/static case will be complicated by the fact
that, in general, many opcodes will appear when scoring 
that were not part of the training set.
While there are several ways to deal with such a situation, 
when scoring, we simply omit any opcodes 
that did not appear in the training set.

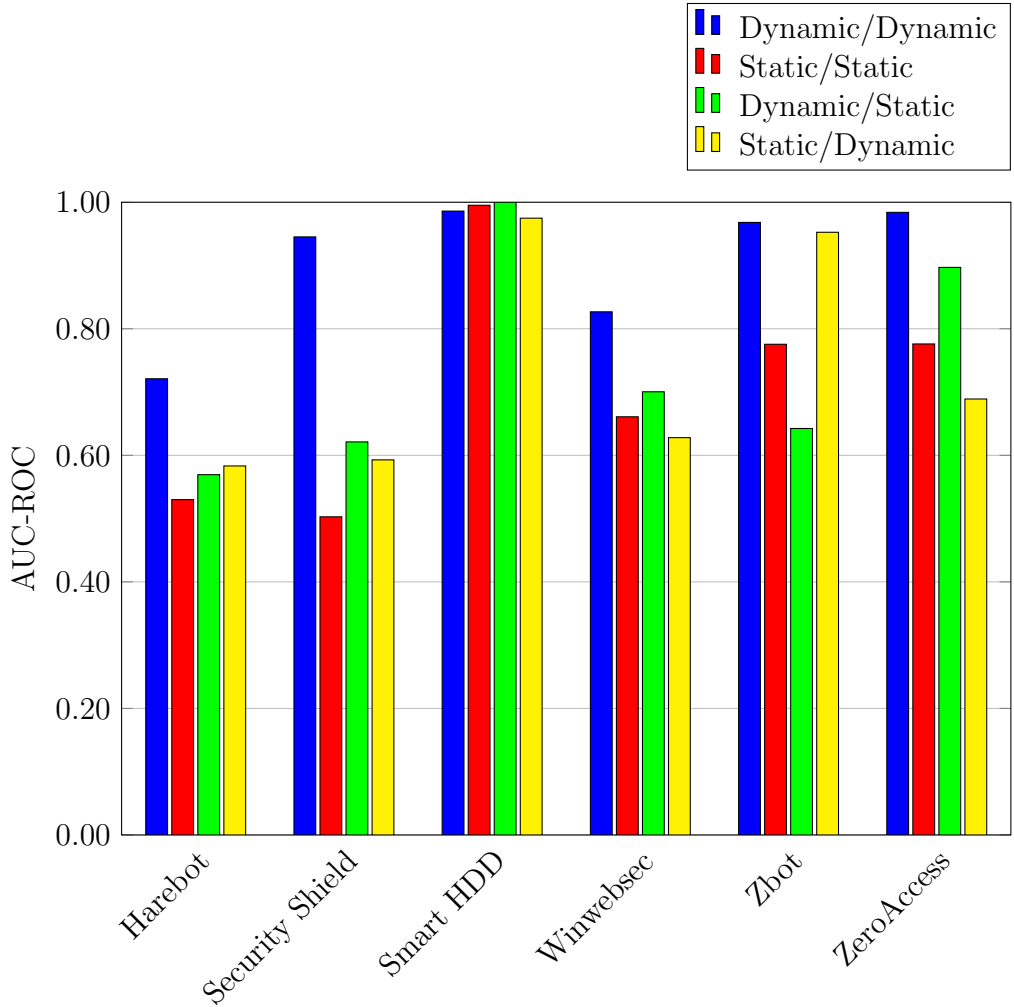
\begin{figure}[htbp]
\begin{center}
\begin{tikzpicture}
    \begin{axis}[
        width  = 0.8*\textwidth,
        height = 10cm,
        ymin=0.0,ymax=1.0,
        major x tick style = transparent,
        ybar=4*\pgflinewidth,
        bar width=8.25pt,
        ymajorgrids = true,
        ylabel = {AUC-ROC},
        symbolic x coords={Harebot,Security Shield,Smart HDD,Winwebsec,Zbot,ZeroAccess},
	y tick label style={
    	/pgf/number format/.cd,
   	fixed,
   	fixed zerofill,
    	precision=2},
        xtick = data,
        x tick label style={rotate=45,anchor=north east, inner sep=0mm},
        scaled y ticks = false,
        enlarge x limits=0.1,
        ymin=0,
        legend cell align=left,
        legend style={
                at={(1,1.05)},
                anchor=south east,
                column sep=1ex
        }
    ]
        \addplot[fill=blue]
            coordinates {
(Harebot,0.7210)
(Security Shield,0.9452)
(Smart HDD,0.9860)
(Winwebsec,0.8268)
(Zbot,0.9681)
(ZeroAccess,0.9840)
};
        \addplot[fill=red]
            coordinates {
(Harebot,0.5300)
(Security Shield,0.5028)
(Smart HDD,0.9952)
(Winwebsec,0.6609)
(Zbot,0.7755)
(ZeroAccess,0.7760)
};
        \addplot[fill=green]
            coordinates {
(Harebot,0.5694)
(Security Shield,0.6212)
(Smart HDD,1.0000)
(Winwebsec,0.7004)
(Zbot,0.6424)
(ZeroAccess,0.8970)
};
        \addplot[fill=yellow]
            coordinates {
(Harebot,0.5832)
(Security Shield,0.5928)
(Smart HDD,0.9748)
(Winwebsec,0.6279)
(Zbot,0.9525)
(ZeroAccess,0.6890)
};
        \legend{Dynamic/Dynamic, Static/Static, Dynamic/Static, Static/Dynamic}
    \end{axis}
\end{tikzpicture}
\end{center}
\vglue-0.3in
\caption{ROC Results for Opcode Sequences\label{AUC_bar_Opcodes}} 
\end{figure}

Our results for training and scoring on opcode sequences 
are given in Table~\ref{auc-opc-reverse}. The results in 
Table~\ref{auc-opc-reverse} are given in the form of a bar graph
in Figure~\ref{AUC_bar_Opcodes}.

\begin{table}[htbp]
\caption{AUC-ROC Results for Opcode Sequences\label{auc-opc-reverse}}
\begin{center}
\def\s{\phantom{/}}
\begin{tabular}{l | cccc}\hline\hline
\multirow{2}{*}{Family} 
& Dynamic/ & Static/ & Dynamic/ & Static/ \\ 
 & Dynamic & Static & Static & Dynamic\\ \hline
Harebot
 & 0.7210
 & 0.5300
 & 0.5694
 & 0.5832
 \\
Security Shield
 & 0.9452
 & 0.5028
 & 0.6212
 & 0.5928
 \\
Smart HDD
 & 0.9860
 & 0.9952
 & 1.0000
 & 0.9748
 \\
Winwebsec
 & 0.8268
 & 0.6609
 & 0.7004
 & 0.6279
 \\
Zbot
 & 0.9681
 & 0.7755
 & 0.6424
 & 0.9525
 \\
ZeroAccess
 & 0.9840
 & 0.7760
 & 0.8970
 & 0.6890
 \\
\hline\hline
\end{tabular}
\end{center}
\end{table}

These opcode-based results are generally not as strong
as those obtained for API call sequences. But, as with API call  
sequences, the best results
are obtained in the dynamic/dynamic case.
However, unlike the API call sequence models, opcode sequences
yield results that are roughly equivalent 
in the static/static and the hybrid dynamic/static case.
Additional experimental results can be found in~\cite{Anusha}.



%

\subsection{Imbalance Problem}

In statistical-based scoring, we typically have a primary test that is used to filter suspect cases,
followed by a secondary test that is applied to these suspect cases. For malware
detection, the primary test is likely to have an imbalance, in the sense that
the number of benign samples exceeds the number of malware samples---possibly
by a large margin. In the secondary stage, we would expect the imbalance to be far 
less significant. Due to their cost, malware detection techniques such as those considered in this paper
would most likely be applied at the secondary stage. Nevertheless, it may be instructive
to consider the effect of a large imbalance between the benign and malware sets.
In this section, we consider the effect of such an imbalance on our
dynamic, static, and hybrid techniques.

We can simulate an imbalance by simply duplicating each benign score~$n$ times.
Assuming that the original number of scored benign and malware samples are equal,
a duplication factor of~$n$ simulates an imbalanced data set where
the benign samples outnumber the malware samples by a factor of~$n$.
Provided that our original benign set is representative, we would expect 
an actual benign set of the appropriate size to yield scores that, on average,
match this simulated (i.e., expanded) benign set.

However, the AUC-ROC for such an expanded benign set will be the same as for the original set. 
To see why this is so,
suppose that for a given threshold, we have~$\TP=a$, $\FN=b$, $\FP=c$, and~$\TN=d$. 
Then~$(x,y)$ is a point on the ROC curve, where
\begin{equation}\label{eq:ROC_pt}
  x =  \FPR = \frac{\FP}{\FP + \TN} = \frac{c}{c + d} \mbox{\ \ and\ \ } y = \TPR = \frac{\TP}{\TP + \FN} = \frac{a}{a + b} 
\end{equation}
Now suppose that we duplicate each element of the negative (i.e., benign) set~$n$ times.
Then for the same threshold used to compute~\eref{eq:ROC_pt},
we have~$\TP=a$, $\FN=b$, $\FP=nc$, and~$\TN=nd$, and hence
we obtain the same point~$(x,y)$ on the ROC curve for this modified dataset.

In contrast, for PR curves, using the same threshold as above we have
$$
  \rec = \frac{\TP}{\TP + \FN}= \frac{a}{a + b} \mbox{\ \ and\ \ } \prec = \frac{\TP}{\TP + \FP} = \frac{a}{a + c}
$$
When we expand our dataset by duplicating the benign scores~$n$ times,
this threshold yields
$$
  \rec = \frac{a}{a + b} \mbox{\ \ and\ \ } \prec = \frac{a}{a + nc}
$$
Consequently, we see that simulating an imbalance in this way will tend to flatten the PR
curve, and thereby reduce the AUC-PR. In addition, the precise degree
of flattening will depend on the relative distribution of the malware
and benign scores.

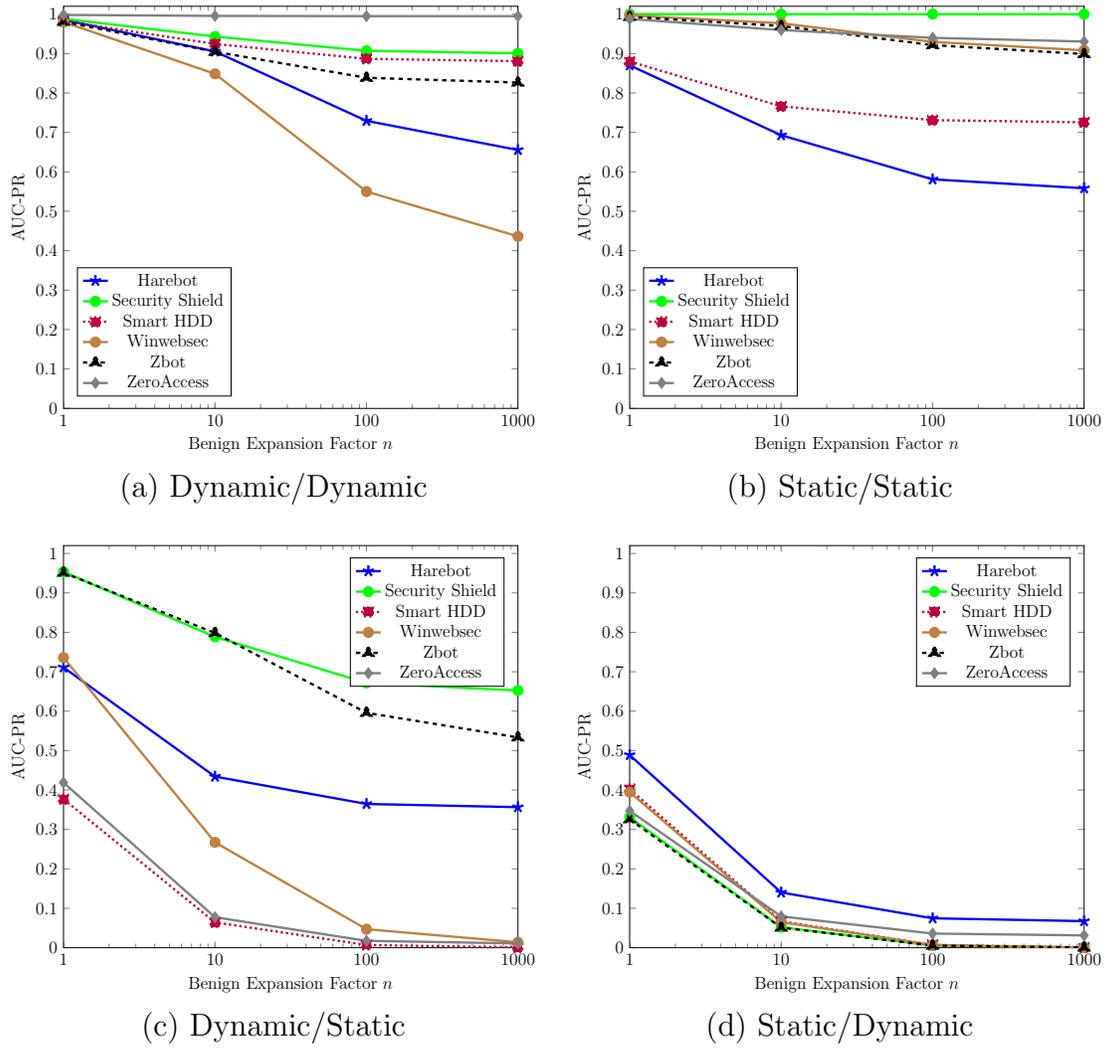
\begin{figure}[htbp]
  \begin{center}
  \begin{tabular}{cc}
    \begin{tikzpicture}[scale=0.55]
    \begin{axis}[width=0.75\textwidth,height=0.675\textwidth,xmin=1.0,xmax=1000.0,
                        /pgf/number format/1000 sep={},
                        xmode=log,
    		       log ticks with fixed point,
                        ymin=0.0,ymax=1.02,legend pos=south west,
                        xlabel={Benign Expansion Factor $n$},ylabel={AUC-PR}] 
        \addplot[color=blue,ultra thick,mark=star,mark size=4.0] coordinates {
(1,0.985773)
(10,0.905363)
(100,0.729423)
(1000,0.655551)};
\addlegendentry{Harebot}
        \addplot[color=green,ultra thick,mark=*,mark size=3.0] coordinates {
(1,0.988363)
(10,0.943171)
(100,0.907063)
(1000,0.900754)};
\addlegendentry{Security Shield}
        \addplot[color=purple,dotted,ultra thick,mark=square*,mark size=3.0] coordinates {
(1,0.982527)
(10,0.924395)
(100,0.886658)
(1000,0.880701)};
\addlegendentry{Smart HDD}
        \addplot[color=brown,ultra thick,mark=otimes*,mark size=3.0] coordinates {
(1,0.980041)
(10,0.848612)
(100,0.550032)
(1000,0.436414)};
\addlegendentry{Winwebsec}
        \addplot[color=black,dashed,ultra thick,mark=triangle*,mark size=4.0] coordinates {
(1,0.980826)
(10,0.904599)
(100,0.838646)
(1000,0.826471)};
\addlegendentry{Zbot}
        \addplot[color=gray,ultra thick,mark=diamond*,mark size=3.0] coordinates {
(1,0.998043)
(10,0.995587)
(100,0.994954)
(1000,0.994880)};
\addlegendentry{ZeroAccess}
    \end{axis}
    \end{tikzpicture}
&
    \begin{tikzpicture}[scale=0.55]
    \begin{axis}[width=0.75\textwidth,height=0.675\textwidth,xmin=1.0,xmax=1000.0,
                        /pgf/number format/1000 sep={},
     	               xmode=log,
    		       log ticks with fixed point,
                        ymin=0.0,ymax=1.02,legend pos=south west,
                        xlabel={Benign Expansion Factor $n$},ylabel={AUC-PR}] 
        \addplot[color=blue,ultra thick,mark=star,mark size=4.0] coordinates {
(1,0.870216)
(10,0.692968)
(100,0.580986)
(1000,0.558366)};
\addlegendentry{Harebot}
        \addplot[color=green,ultra thick,mark=*,mark size=3.0] coordinates {
(1,1.0000)
(10,1.0000)
(100,1.0000)
(1000,1.0000)};
\addlegendentry{Security Shield}
        \addplot[color=purple,dotted,ultra thick,mark=square*,mark size=3.0] coordinates {
(1,0.879947)
(10,0.766204)
(100,0.731166)
(1000,0.725662)};
\addlegendentry{Smart HDD}
        \addplot[color=brown,ultra thick,mark=otimes*,mark size=3.0] coordinates {
(1,0.996732)
(10,0.976936)
(100,0.929683)
(1000,0.908319)};
\addlegendentry{Winwebsec}
        \addplot[color=black,dashed,ultra thick,mark=triangle*,mark size=4.0] coordinates {
(1,0.993100)
(10,0.970000)
(100,0.921559)
(1000,0.898880)};
\addlegendentry{Zbot}
        \addplot[color=gray,ultra thick,mark=diamond*,mark size=3.0] coordinates {
(1,0.987925)
(10,0.959666)
(100,0.939926)
(1000,0.930498)};
\addlegendentry{ZeroAccess}
    \end{axis}
    \end{tikzpicture}
\\
(a) Dynamic/Dynamic & (b) Static/Static \\[2.5ex]
    \begin{tikzpicture}[scale=0.55]
    \begin{axis}[width=0.75\textwidth,height=0.675\textwidth,xmin=1.0,xmax=1000.0,
                        /pgf/number format/1000 sep={},
     	               xmode=log,
    		       log ticks with fixed point,
                        ymin=0.0,ymax=1.02,legend pos=north east,
                        xlabel={Benign Expansion Factor $n$},ylabel={AUC-PR}] 
        \addplot[color=blue,ultra thick,mark=star,mark size=4.0] coordinates {
(1,0.711127)
(10,0.434067)
(100,0.364677)
(1000,0.356483)};
\addlegendentry{Harebot}
        \addplot[color=green,ultra thick,mark=*,mark size=3.0] coordinates {
(1,0.953410)
(10,0.788649)
(100,0.671651)
(1000,0.652295)};
\addlegendentry{Security Shield}
        \addplot[color=purple,dotted,ultra thick,mark=square*,mark size=3.0] coordinates {
(1,0.376798)
(10,0.064366)
(100,0.006981)
(1000,0.000704)};
\addlegendentry{Smart HDD}
        \addplot[color=brown,ultra thick,mark=otimes*,mark size=3.0] coordinates {
(1,0.735914)
(10,0.267325)
(100,0.047021)
(1000,0.013888)};
\addlegendentry{Winwebsec}
        \addplot[color=black,dashed,ultra thick,mark=triangle*,mark size=4.0] coordinates {
(1,0.951260)
(10,0.798286)
(100,0.595811)
(1000,0.533579)};
\addlegendentry{Zbot}
        \addplot[color=gray,ultra thick,mark=diamond*,mark size=3.0] coordinates {
(1,0.418968)
(10,0.077417)
(100,0.017452)
(1000,0.010981)};
\addlegendentry{ZeroAccess}
    \end{axis}
    \end{tikzpicture}
&
    \begin{tikzpicture}[scale=0.55]
    \begin{axis}[width=0.75\textwidth,height=0.675\textwidth,xmin=1.0,xmax=1000.0,
                        /pgf/number format/1000 sep={},
     	               xmode=log,
    		       log ticks with fixed point,
                        ymin=0.0,ymax=1.02,legend pos=north east,
                        xlabel={Benign Expansion Factor $n$},ylabel={AUC-PR}] 
        \addplot[color=blue,ultra thick,mark=star,mark size=4.0] coordinates {
(1,0.488821)
(10,0.140107)
(100,0.074617)
(1000,0.067468)};
\addlegendentry{Harebot}
        \addplot[color=green,ultra thick,mark=*,mark size=3.0] coordinates {
(1,0.331180)
(10,0.051991)
(100,0.005530)
(1000,0.000557)};
\addlegendentry{Security Shield}
        \addplot[color=purple,dotted,ultra thick,mark=square*,mark size=3.0] coordinates {
(1,0.402514)
(10,0.066878)
(100,0.007192)
(1000,0.000725)};
\addlegendentry{Smart HDD}
        \addplot[color=brown,ultra thick,mark=otimes*,mark size=3.0] coordinates {
(1,0.394676)
(10,0.064962)
(100,0.006966)
(1000,0.000702)};
\addlegendentry{Winwebsec}
        \addplot[color=black,dashed,ultra thick,mark=triangle*,mark size=4.0] coordinates {
(1,0.325950)
(10,0.051075)
(100,0.005434)
(1000,0.000547)};
\addlegendentry{Zbot}
        \addplot[color=gray,ultra thick,mark=diamond*,mark size=3.0] coordinates {
(1,0.347228)
(10,0.079469)
(100,0.035932)
(1000,0.031289)};
\addlegendentry{ZeroAccess}
    \end{axis}
    \end{tikzpicture}
\\
(c) Dynamic/Static & (d) Static/Dynamic \\[1.5ex]
\end{tabular}
  \end{center}
  \vglue -0.20in
  \caption{AUC-PR and Imbalanced Data (API Calls)\label{fig:AUC_PR_API_graphs}}
\end{figure}

We have shown that the AUC-ROC provides no information on the effect of an imbalance
between the malware and benign sets. In some sense, this can be viewed as a strength
of the AUC-ROC statistic, although it does render it useless for analyzing the effect of imbalanced data.
On the other hand, the AUC-PR is a useful statistic for comparing the effect of
an imbalance between these two sets. Consequently, we use the AUC-PR
in this section to determine the effect of a (simulated) imbalance between the
malware and benign sets. We consider the API sequence results,
and we duplicate each benign score
by a factor of~$n=1$, $n=10$, $n=100$, and~$n=1000$,
and plot the results on a logarithmic (base 10) scale. The
resulting AUC-PR values for each of the four
cases (i.e., dynamic/dynamic, static/static, dynamic/static,
and static/dynamic) are plotted as line graphs in Figure~\ref{fig:AUC_PR_API_graphs}.

The results in Figure~\ref{fig:AUC_PR_API_graphs} suggest that we
can expect the superiority
of the a fully dynamic approach,
to increase as the imbalance between the benign and malware sets grows.
In addition, the advantage of the fully static approach over our
hybrid approaches increases as the imbalance increases.
We also see that even in those cases where the dynamic/static approach is initially 
competitive, it fails to remain so for a large imbalance. And finally, the 
overall weakness of the static/dynamic approach is even more apparent 
from this PR analysis.

\subsection{Discussion}

The results in this section show that for API calls and opcode sequences,
a fully dynamic strategy is generally the most effective approach.
However, dynamic analysis is generally costly in comparison to static analysis. 
At the training phase, this added cost is not a significant issue, since
training is essentially one-time work that can be done offline.
But, at the scoring phase, dynamic analysis would likely be impractical,
particularly where it is necessary to scan a large number of files.

In a hybrid approach, we might attempt to improve the training phase by using 
dynamic analysis while, for the sake of efficiency, using only a static approach
in the scoring phase. However, such a strategy was not particularly 
successful in the experiments considered here.
For API call sequences, we consistently obtained worse results with
the hybrid dynamic/static as compared to a fully static approach. For opcode
sequences, the results were inconsistent---in four of the cases,
the hybrid dynamic/static method was marginally better than the fully 
static approach, but for one case it was significantly worse.

Attempting to optimize a malware detection technique by
using hybrid analysis is intuitively appealing. 
While such a hybrid approach may be more effective in certain cases, 
our results show that this is not likely to be generically the case.
Consequently, when hybrid approaches are proposed, it
would be advisable to test the results against comparable
fully dynamic and fully static techniques.

\section{Conclusion and Future Work\label{sect:conclusion}}

In this paper, we tested malware detection techniques
based on API call sequences and opcode sequences.
We trained Hidden Markov Models and compared detection
rates for models based on static data, dynamic data, and
hybrid approaches. 

Our results indicate that a fully dynamic approach based on
API calls is extremely effective across a range of malware
families. A fully static approach based on API calls was nearly
as effective in most cases.
Our results also show that opcode sequences can be effective in
many cases, but for some families the results are not impressive.
These results likely reflect the nature of obfuscation techniques
employed by malware writers. That is, current obfuscation techniques
are likely to have a significant effect on opcode sequences,
but little attention is paid to API calls. With some additional 
effort, API call sequences could likely be obfuscated, in which case the 
advantage of relying on API call sequences for detection
might diminish significantly. 

Examples of relatively complex and involved
hybrid techniques have recently appeared in the literature.
However, due to the use of different data sets, different measures of
success, and so on, it is often difficult, if not impossible, 
to compare these techniques to previous (non-hybrid) work.
Further, the very complexity of such detection techniques often makes it
difficult to discern the actual benefit of any one particular aspect of 
a technique. The primary goal of this research was to test the tradeoffs between
static, dynamic, and hybrid analysis, while eliminating as many 
confounding variables as possible.

The experimental results presented in this paper indicate that
a straightforward hybrid approach is unlikely to be 
superior to fully dynamic detection. 
And even in comparison to fully static detection,
our hybrid dynamic/static approach did not offer consistent improvement.
Interestingly, the impractical static/dynamic hybrid approach
was superior in some cases (by some measures).
These results are, perhaps, somewhat surprising given the claims 
made for hybrid approaches. 

Of course, it is certain that hybrid techniques offer significant benefits in some cases.
But, the work here suggests that such claims should be subject to
careful scrutiny. In particular, it should be made clear whether
improved detection is actually due to a hybrid model itself,
or some other factor, such as the particular combination of scores used. 
Furthermore, it should be determined
whether these benefits exist over a wide range of malware
samples, or whether they are only relevant for a relatively narrow
range of malware.

Future work could include a similar analysis invovling additional features
beyond API calls and opcodes. A comparison of scoring
techniques other than HMMs (e.g., graph-based scores, structural
scores, other machine learning and statistical scores) and optimal
combinations of static and dynamic scores (e.g., using Support Vector Machines)
would be worthwhile. Finally, a more in-depth analysis of imbalance
issues in this context might prove interesting.

%
%


\end{document}